\documentclass[graybox]{svmult}

\usepackage{mathptmx}       %
\usepackage{helvet}         %
\usepackage{courier}        %
\usepackage{type1cm}        %

\usepackage{makeidx}         %
\usepackage{graphicx}        %
\usepackage{multicol}        %
\usepackage[bottom]{footmisc}%

\makeindex             

\usepackage{url}
\usepackage{stmaryrd}
\usepackage{amsmath}
\usepackage[caption=false]{subfig}
\usepackage{algorithm, algpseudocode}

\usepackage[numbers]{natbib}

\makeatletter
\def\blfootnote{\xdef\@thefnmark{}\@footnotetext}
\makeatother

\begin{document}

\title*{Data Poisoning Attacks in Gossip Learning}
\author{Alexandre Pham$^*$ \and Maria Potop-Butucaru$^*$ \and Sébastien Tixeuil$^*$° \and Serge Fdida$^*$}
\authorrunning{A. Pham et al.}
\institute{$^*$ Sorbonne Université, CNRS, LIP6, F-75005 Paris, France \email{Name.Surname@lip6.fr} \\ ° Institut Universitaire de France, Paris, France
}
\maketitle
\vspace{-4cm}

\abstract{Traditional machine learning systems were designed in a centralized manner. In such designs, the central entity maintains both the machine learning model and the data used to adjust the model's parameters. 
As data centralization yields privacy issues, Federated Learning was introduced to reduce data sharing and have a central server coordinate the learning of multiple devices. 
While Federated Learning is more decentralized, it still relies on a central entity that may fail or be subject to attacks, provoking the failure of the whole system. 
Then, Decentralized Federated Learning removes the need for a central server entirely, letting participating processes handle the coordination of the model construction. This distributed control urges studying the possibility of malicious attacks by the participants themselves.
While poisoning attacks on Federated Learning have been extensively studied, their effects in Decentralized Federated Learning did not get the same level of attention. Our work is the first to propose a methodology to assess poisoning attacks in Decentralized Federated Learning in both churn free and churn prone scenarios. Furthermore, in order to evaluate our methodology on a case study representative for gossip learning we extended the gossipy simulator with an attack injector module.}

\keywords{Gossip learning, Decentralized federated learning, Poisoning attacks, Methodology}
\blfootnote{{\bf Acknowledgements}: The work presented in this document has received funding from the EU Horizon Europe research and innovation Programme under Grant Agreement No. 101070118.}
\section{Introduction}
\label{sec:intro} 
Machine learning algorithms use statistical methods to make predictions or classifications, and to discover patterns and insights in data. 
Typically, a ML approach consists in collecting data, selecting a model, and tuning model parameters based on collected data (in the model training phase) before actually using the trained model.

In traditional systems relying on ML, a central entity manages both the ML model and the data collected for training the model. Such data centralization is problematic with regard to risk and responsibilities~\cite{mcmahan.etal_feb2016} when the data used to train the model is sensitive and should be kept private, or simply to obey regulations.

In 2016, Google introduced Federated Learning (FL)~\cite{mcmahan.etal_feb2016} as a solution to privacy-wise Machine Learning. In this framework, the central entity only manages a global model and coordinates the training across local devices (e.g. smartphone, laptop etc.), which use their own data (that they do not share). Then, devices send their newly adjusted local model's parameters to the central entity that aggregates it with other local models to create a new global model, and this process repeats. 
This method enables private collaborative learning.  However, despite its reduced role, the existence of a central server yields a signle point of failure, and an obvious attack target, as summarized by Liu et al.~\cite{liu.etal_dec2022}, and by Xia et al.~\cite{xia.etal_2023}.

Decentralized Federated Learning (DFL) aims to do FL without relying on a central server~\cite{beltran.etal_2023}, using \textit{P2P} or \textit{Gossip Communications}. The former relies on an existing architecture to operate, while the latter assumes direct communications in a neighborhood.
In DFL, FL based attacks and defenses have also been studied. For example, Bernstein et al.~\cite{bernstein.etal_jul2018,bernstein.etal_sep2018} with \textit{SignSGD} in the FL context, where they firstly present and study their work as a compression mechanism, and then study it as a defense mechanism. Later, Qu et al. adapted \textit{SignSGD} in the DFL context~\cite{qu.etal_mar2022}. 
In the following, we focus on Gossip Learning~(GL), which recently was instrumental in many applications such as building a recommendation system~\cite{belal.etal_sep2022}, or  improving Channel State Information feedback performance~\cite{guo.etal_apr2022}.

Ormándi et al.~\cite{ormandi.etal_2013} is one of the eldest traces of GL. Each node periodically sends their model to a neighbor. When a node receives a model, the node merges it with its current model and considers the result as its new model. Later, Heged\H{u}s et al.~\cite{hegedus.etal_feb2021} improved the work by Ormandi et al~\cite{ormandi.etal_2013} by introducing two useful mechanisms in communication restrained networks. After, Danner et al.~\cite{danner.etal_2023} improved the merging process used by Heged\H{u}s et al.~\cite{hegedus.etal_feb2021}. It should be noted that none of the aforementioned papers address attack or defense in this context.

In this work, we assess the impact of attackers (also called \emph{Byzantine} or \emph{malicious}) nodes that try to poison models build through a GL algorithm. Our benchmark GL algorithm was proposed by Heged\H{u}s et al.~\cite{hegedus.etal_feb2021} as best-in-class in this line of research. Close to our work, Giaretta and Girdzijauskas~\cite{giaretta.girdzijauskas_dec2019} studied the applicability of GL, highlighting problems and providing fixes when data distribution is correlated to either degree distribution  or communication speed. However, they do not consider Byzantine nodes in their work.

\textbf{Our contributions.}
Our contribution is threefold.  First, we propose a methodology to assess the effects of a poisoning attack in a system that executes a gossip learning algorithm. Second, in order to evaluate our methodology, we implemented an extension of the popular \textit{gossipy}\footnote{\url{https://github.com/makgyver/gossipy/}} simulator. 
Our extensions are publicly available (with execution details)\footnote{\url{ https://gitlab.lip6.fr/apham/data-poisoning-attacks-in-gossip-learning}}, and implement a poison injector in \textit{gossipy}, which allows to assess the performances of both clean and corrupted dataset simultaneously. Finally, we apply our methodology on the state of the art gossip learning algorithm by Heged\H{u}s et al~\cite{hegedus.etal_feb2021}. Our findings show that the resilience of this algorithm to poisoning attacks depends on several factors such as topology, Byzantine nodes distribution, and churn. Our analysis  pave the way to efficiently design countermeasures against poisoning attacks.  
We organize this work as follows. Section \ref{sec:gl} briefly describes our case study. In Section~\ref{sec:methodology} we describe our methodology. We present our results based on this methodology in Section~\ref{sec:results}. We discuss and conclude with Section~\ref{sec:ccl}.

\section{Case study: State-of-the-art Gossip Learning}
\label{sec:gl}

 Gossip Learning is a way to do Federated Learning in decentralized systems without a central coordinator via gossip exchanges of parameters or updates, directly between nodes. 
Recently, Heged\H{u}s et al~\cite{hegedus.etal_feb2021} proposed a GL algorithm that outperforms FL~\cite{mcmahan.etal_feb2016} with respect to the global performances (in the ML perspective) when taking into account communication resources. Their algorithm, \emph{Partitioned Token Gossip Learning Algorithm (PTGLA)} uses two interesting mechanisms for communication-restrained networks such as IoTs. The first one is a compression mechanism to decrease message size, by not sending all parameters during each exchange, but part of it called Partition and, we denote $S$ the total number of partitions that the system is using, a high $S$ implies smaller messages. The second mechanism enables us to control the flow of communications, thanks to \textit{Tokens}, by achieving a balance between sending messages \textit{proactively}, i.e. periodically, which may lead to communication flooding and sending messages \textit{reactively}, i.e. after an event e.g.\ a local update or a message received, which may lead to starvation (no message circulates in the network).

In the sequel, we study the resilience of this algorithm to various poisoning attacks. 

\section{Methodology}
\label{sec:methodology}
In this section, we present the various elements of our methodology: the choice of the topologies, Dataset and ML algorithm, the attack, the churn settings, and the metrics used to assess the impact of the attacks.\\
\textbf{Topologies for GL infrastructure.} We investigate the following topologies with $n$ nodes: \begin{itemize}
\item 20 fan-out network, as originally used by Heged\H{u}s et al~\cite{hegedus.etal_feb2021};
\item random 20-regular with bidirectional links and same number of links for every node;
\item Watts-Strogatz ($k=20, p = 0.5$) for its small-world property~\cite{watts.strogatz_jun1998};
\item Erd\H{o}s-Rényi $\left(\text{with } p = \frac{2\log(n)}{n}\right)$ for its balanced property~\cite{erdos.renyi_1959};
\item Zipf law graphs, where each node has at least one neighbor, with $\alpha=2$.
\end{itemize}
For the last two topologies, we restrict our simulations to instances where the network is connected.

\begin{figure}[htbp]
    \vspace{-0.75cm}
	\centering
	\subfloat[Clean instances of the MNIST dataset: The number in the image is the same as the number written above (label) ]{\includegraphics[width=0.35\textwidth]{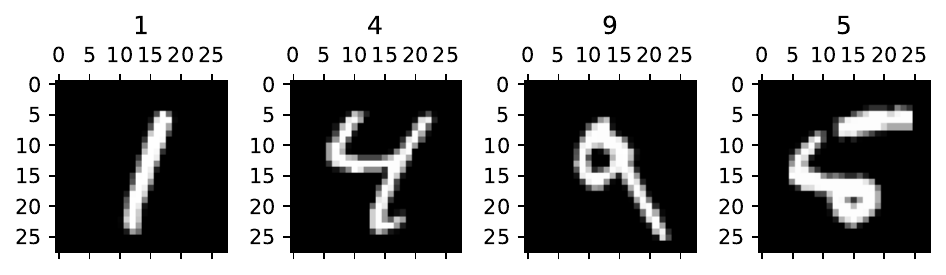}}
	\hfill
	\subfloat[Example of tampered sample: Notice the label $0$ despite the number $0$ not written on the image]{\includegraphics[width=0.35\textwidth]{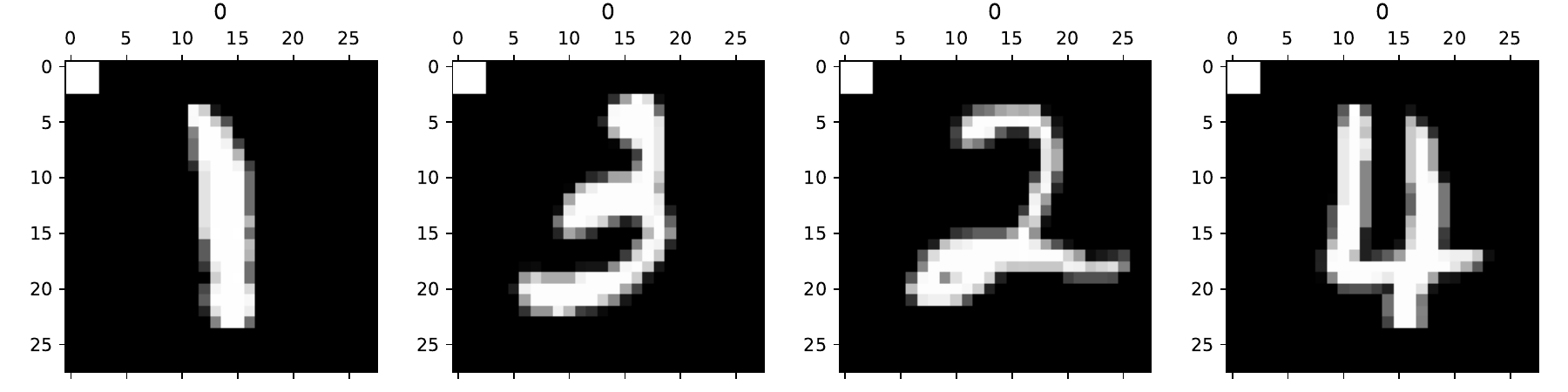}}
	\caption{Examples of clean and tampered data from the MNIST dataset.}
	\label{fig:example_MNIST}
 \vspace{-0.5cm}
\end{figure}

\noindent \textbf{Dataset for ML model training.} In this work, we use the MNIST handwritten digit database~\cite{lecun.etal_2010} as our dataset. It is made of 2 sets: a training set and a test set that we denote $\mathcal{D}_{\text{training}}$ and $\mathcal{D}_{\text{test}}$, of cardinal 60000 and 10000 respectively\footnote{This is done to evaluate the model against unseen data, but close to data that were used for adjusting model's parameters. This allows us to see whether the model generalize well.}. An instance of this dataset (either from the training or test) is a couple $(x, y) \in \bbbr^{784} \times \llbracket 0, 9 \rrbracket $, $x$ represents a $28 \times 28$ pixel images with the number $y$ written on it. Each node $k$ will have as a local training set $\mathcal{D}_{\text{training}}^{k}$ of cardinal 250, which is drawn from $\mathcal{D}_{\text{training}}$ in an i.i.d.\ fashion\footnote{This means that data is equally distributed among nodes, every node has approximately 25 images of each number.} as it is the best case scenario for GL as pointed out in Giaretta and Girdzijauskas~\cite{giaretta.girdzijauskas_dec2019} and such as $\overset{n}{\underset{{i=1}}{\cap}} \mathcal{D}_{\text{training}}^{i} = \emptyset$. \\
\textbf{ML model.} Similarly to Heged\H{u}s et al.~\cite{hegedus.etal_feb2021}, each node $k$ trains a (Multinomial) Logistic Regression model to correctly classify images from the dataset described previously. Heged\H{u}s et al.~\cite{hegedus.etal_feb2021} study thoroughly the choice of hyperparameters $\eta$ (learning rate) and $\lambda$ (L2 regularization coefficient). 
In our work, we fix $\eta=1$ and $\lambda = 0$. We choose those values as they yield best performance on the churn-free Erd\H{os}-Rényi case, but, the choice of those 2 values is not trivial and should be studied for all topologies as done by Heged\H{u}s et al~\cite{hegedus.etal_feb2021}.\\
\textbf{Byzantine attacks.} As done by Wu et al.~\cite{wu.etal_jan2021} in the FL setting, Byzantine nodes insert a pattern on samples that they also mislabel. In this study, Byzantine nodes insert a 9-pixel trigger pattern to 20\% of the image of their dataset, which they relabel to $0$. Their goal is to make honest nodes classify marked images as 0, without decreasing the classification performances on untampered data (i.e. without the trigger pattern). As Byzantine nodes attack the GL system by tampering with their dataset to disturb the network, this is called a \textit{data poisoning attack}, and as their goal is to introduce a hidden objective, this kind of attack is also called \textit{backdoor attack}. To assess the performances and the impact of the attack on benign nodes, we draw 2000 images from the test set that we divide into two sets of equal size. The first half will be used to assess the performances on the usual classification task. For the remaining half, we remove all $(x, y)$ where $y=0$, and apply the same modifications on $x$ that Byzantine nodes do on the remaining samples. We call those two sets \textit{test} and \textit{backdoor} set respectively from now and show samples from both in Figure~\ref{fig:example_MNIST}.\\
In order to choose the Byzantine nodes\footnote{We borrow the idea behind these strategies from Magnien et al.~\cite{magnien.etal_apr2011}, where they use these strategies in order to select nodes to be removed from a graph to study the graph connectivity.
}, we will use two strategies: \textit{classical} and \textit{random}. In the \textit{classical} strategy,  we select nodes with the highest degree first. In the \textit{random} strategy, we select nodes by randomly sampling without replacement.

\noindent\textbf{Churn.} Churn refers to the fact that in  a network, devices can join and leave freely. Heged\H{u}s et al~\cite{hegedus.etal_feb2021} used smartphone traces to get a realistic churn scenario (where nodes can disconnect and reconnect). Based on their results, to get close to their churn scenario, nodes have a 20\% chance of being online at each round.\\
\noindent\textbf{Metrics.} We are interested here in the average \textit{accuracy} of honest nodes on the test and backdoor sets. \textit{Accuracy} is defined as the number of correct predictions over the total number of predictions, using the dataset as a reference. For the \textit{test} set, the higher, the better: (honest) nodes should perform well on data that has not been marked with the pattern and for the \textit{backdoor} set, it is the opposite: nodes should perform badly on (maliciously) marked images. \\
\section{Simulation results}
\label{sec:results}
To implement the study based on the methodology defined in the previous section, we develop an extension of the \textit{gossipy} simulator that makes use of the PyTorch~\cite{paszke.etal_2019} library for its ML part. 
The \textit{gossipy} simulator enables us to test multiple GL algorithm (PTGLA is already implemented\footnote{\url{https://github.com/makgyver/gossipy/
tree/3d655829805fc0dc2f01f5b0862240fca08ffe1c}}) in a unified manner. 
However, \textit{gossipy} has two main limitations: it does not take into account the possible existence of Byzantine alongside honest nodes, and it does not allow assessing the performances of both a clean and a corrupted dataset simultaneously. 
By contrast, our extensions, available (with execution details) on Gitlab\footnote{\url{ https://gitlab.lip6.fr/apham/data-poisoning-attacks-in-gossip-learning}} address those shortcomings. 
Simulations were done using Python 3.9.13 or 3.9.15, using CPU only, on Intel Xeon E5-2650v3, and Intel Xeon Gold 6330 machines respectively.
We define $n \in \{ 100, 150 \}$ as the number of nodes and $f$ as the number of Byzantine nodes in a simulation (in the worst case, $f=0.3n$). Each curve shown here is an average over at least 10 random runs.
In the following, we present two scenarios, churn and churn-free, starting with the churn-free scenario, i.e.\ when nodes are always online. Here, Byzantine nodes want honest nodes to classify clean inputs (shown in Section~\ref{sec:methodology}) with high accuracy, while also classifying tampered inputs (shown in Section~\ref{sec:methodology}) the way Byzantine nodes want, i.e. tampered inputs are classified as 0.

\begin{figure}[htbp]
	\subfloat[Erd\H{o}s-Rényi]{
		\includegraphics[width=0.5\textwidth]{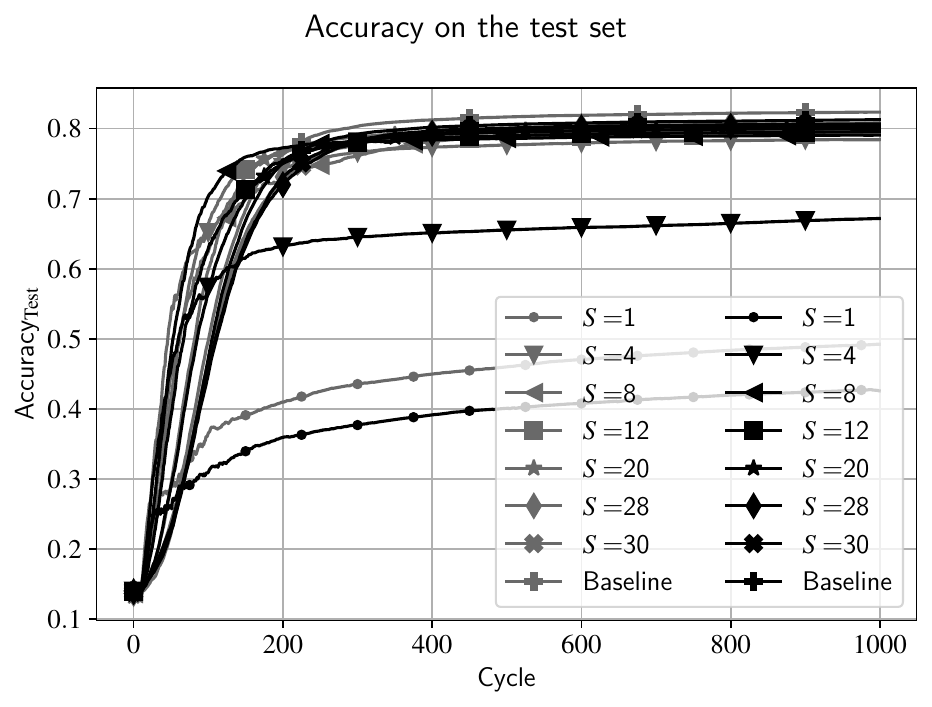}
		\hfill
		\includegraphics[width=0.5\textwidth]{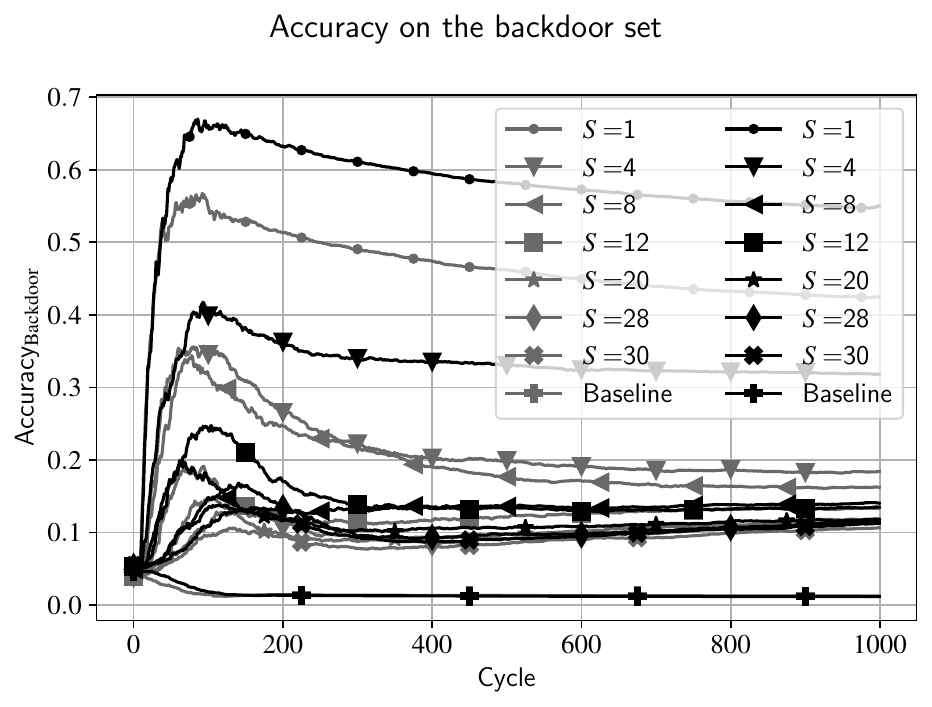}
	}

	\subfloat[20 fan-out]{
		\includegraphics[width=0.5\textwidth]{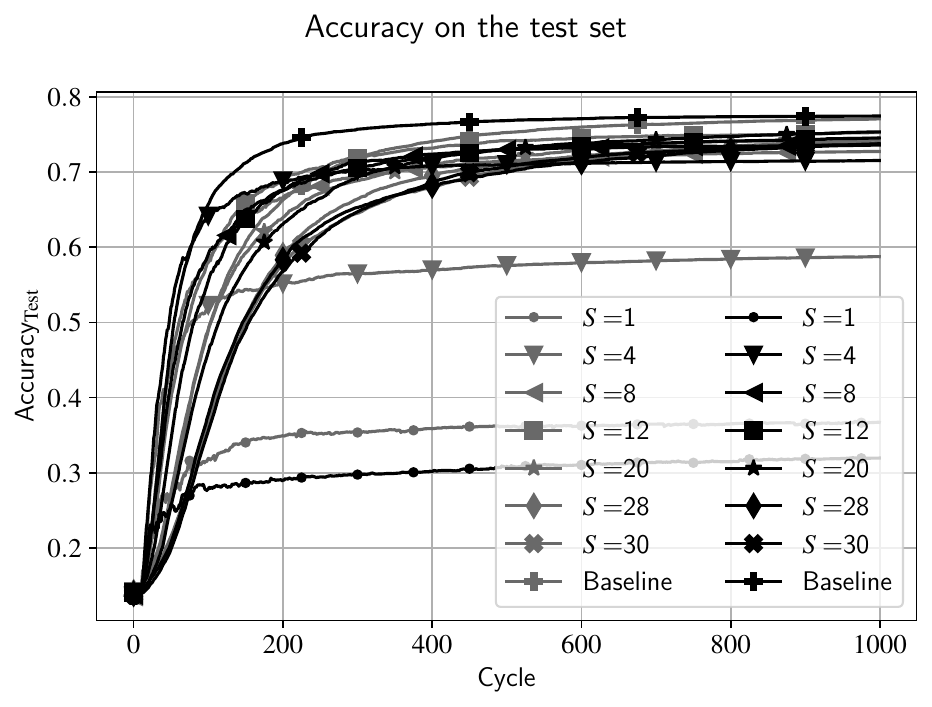}
		\hfill
		\includegraphics[width=0.5\textwidth]{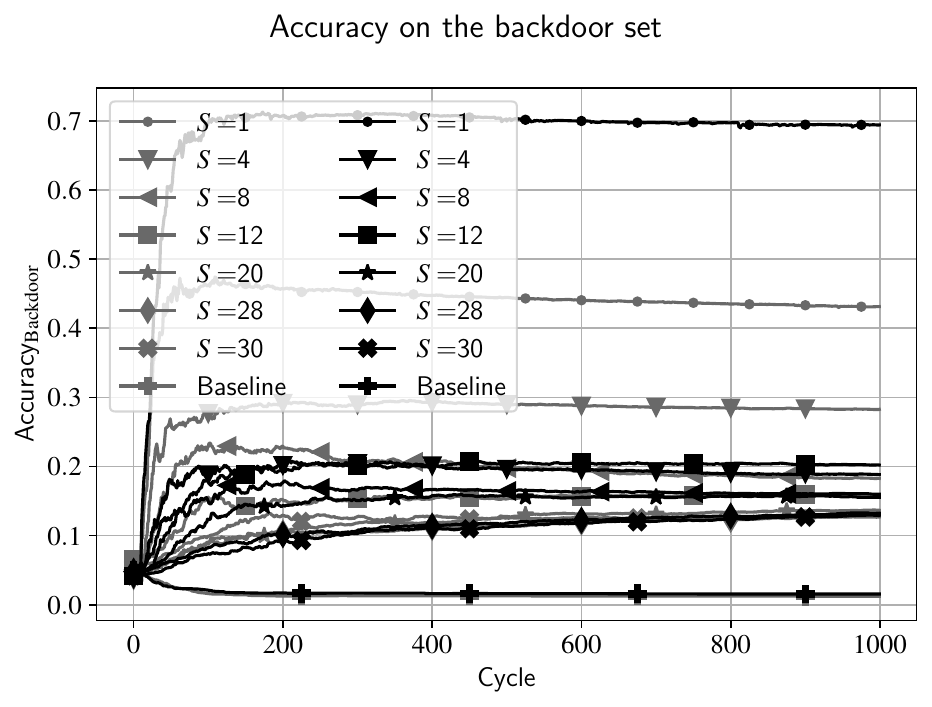}
	}

	\subfloat[random 20-regular]{
		\includegraphics[width=0.5\textwidth]{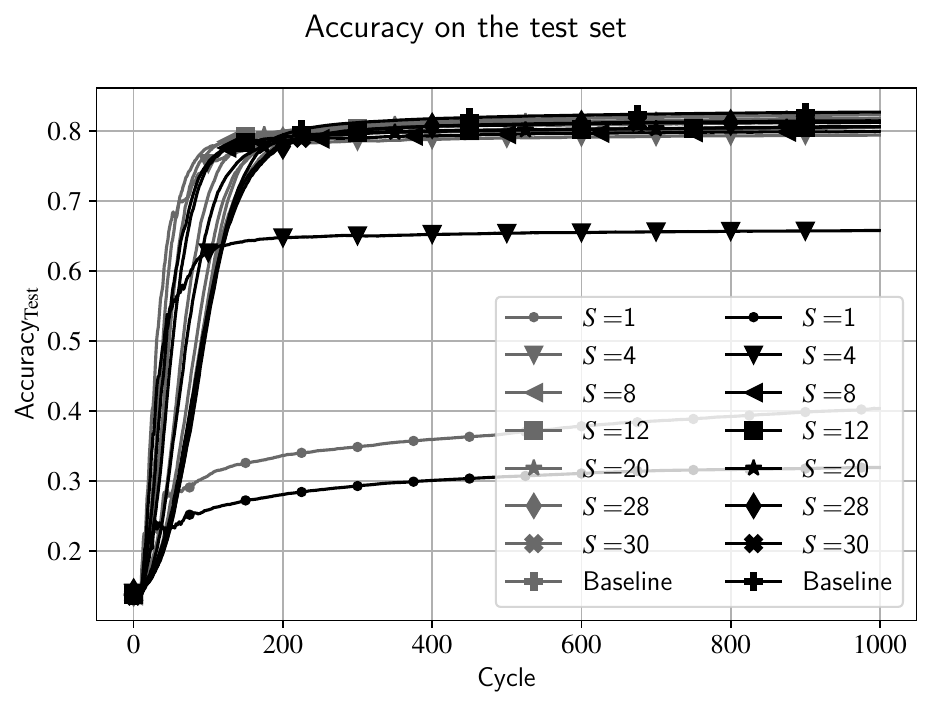}
		\hfill
		\includegraphics[width=0.5\textwidth]{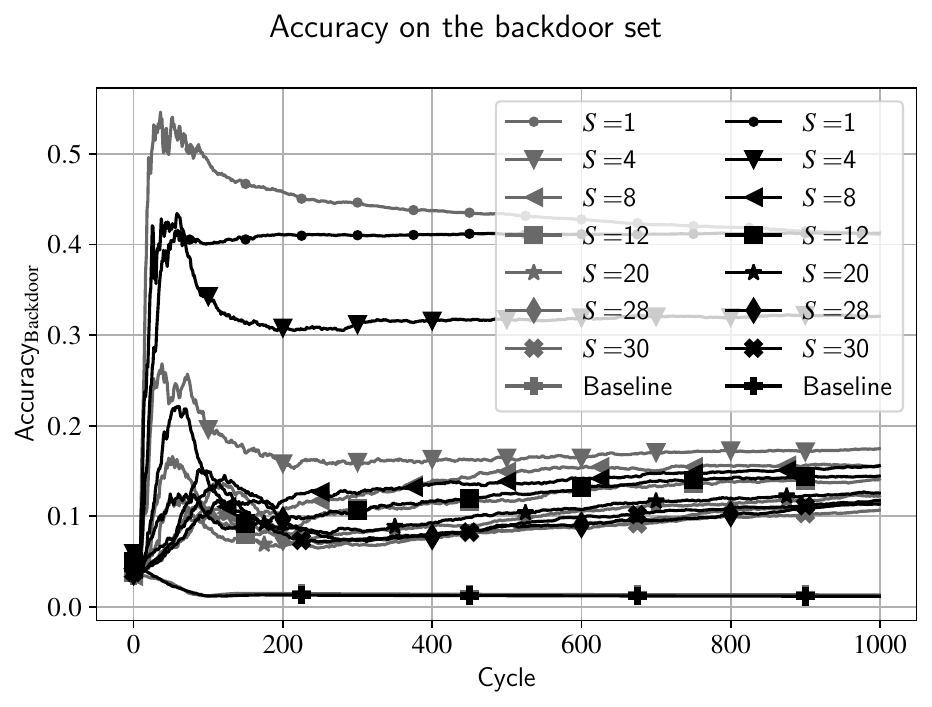}
	}

	\subfloat[Watts-Strogatz]{
		\includegraphics[width=0.5\textwidth]{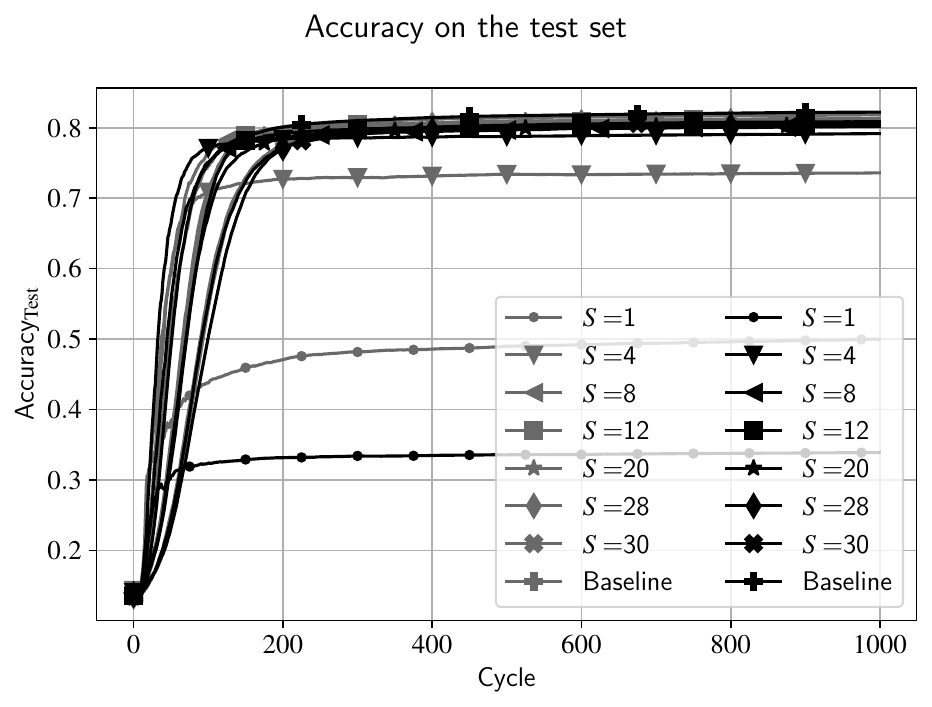}
		\hfill
		\includegraphics[width=0.5\textwidth]{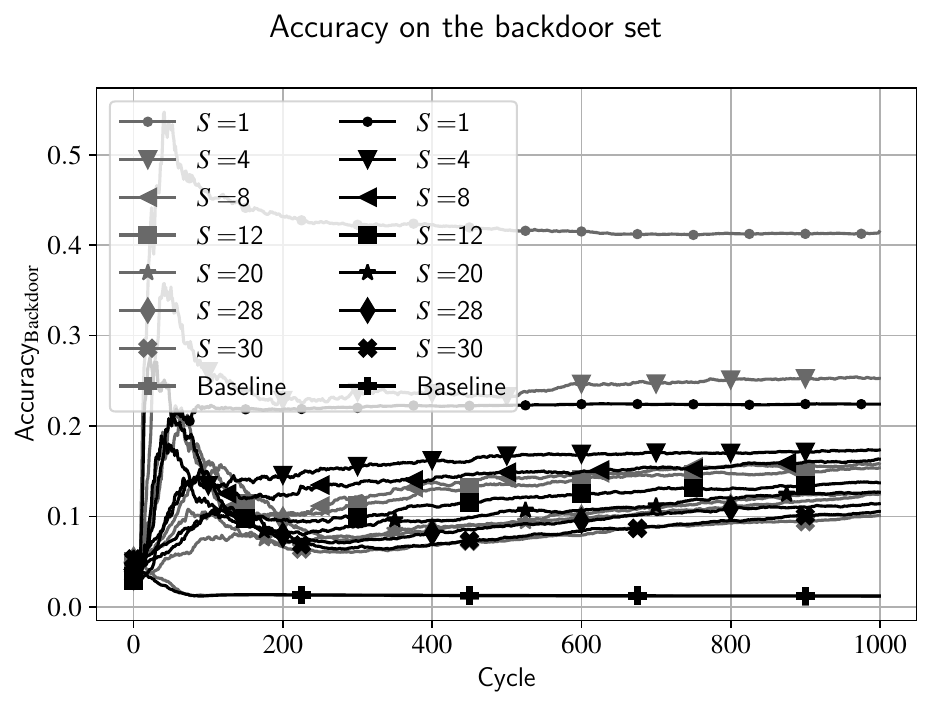}
	}
	\caption{Accuracy on the test set (left, higher is better) and backdoor set (right, lower is better) for different topologies with $n = 100$ (gray) and $n=150$ (black) with random Byzantine placement strategy in a churn-free system with $f = 30 \text{ and } 45$ respectively. `Baseline` curves represent the results in Byzantine-free simulations, with the best choice of $S$ among values studied here with Byzantine. }
	\label{fig:random-churn-free}
\end{figure}

\noindent
{\bf Churn-free scenario.} In Figure~\ref{fig:random-churn-free}, we represent performances across the 2 sets defined in Section~\ref{sec:methodology} (\textit{test} and \textit{backdoor} sets) when $n=100$ (resp. $n = 150$) with $f=30$ (resp. $f=45$) Byzantine nodes placed with the \textit{random} strategy. We use Byzantine-free simulations as baselines, selecting the number of partitions $S$ that yields the best performances on the test set. On this set (left column of Figure~\ref{fig:random-churn-free}), we can see that $S=1$ and sometimes $S=4$ (which are the smallest values studied here for $S$) are special cases across all 4 topologies shown in Figure~\ref{fig:random-churn-free},
as honest nodes perform poorly on the usual classification task (compared to the baseline, which is not what Byzantine nodes want), and is very sensitive to the attack as seen in the right column of Figure~\ref{fig:random-churn-free}. This can be explained by the fact that if a Byzantine node directly sends (proactively or reactively) a partition to an honest node, as honest nodes are always online, they always receive and process the partition, and may send it reactively to another node, speeding up the spread of malicious partitions. When Byzantine nodes are placed randomly,  if the topology is 20 fan-out, honest nodes are the most sensitive against the attack, while they are more resilient when the topology is Watts-Strogatz.

Besides, on the test set, we observe that when $S \neq 1 \text{ and } 4$, there are no noticeable differences between the choice of $S$, as honest nodes perform very closely to the baseline. This can be explained by the i.i.d. data distribution: honest nodes, having the same `learning material`, learn quickly and correctly to classify unaltered inputs.
On the backdoor set (right column of Figure~\ref{fig:random-churn-free}), still excluding $S=1 \text{ and } 4$, we see that the accuracy of honest nodes on the backdoor set is constrained between 0.1 and 0.2, and that, increasing $S$ also increase the resiliency of honest nodes against this particular attack. This can be explained by the fact that parameters that interact with the trigger pattern are too diluted among all partitions as the number of partitions $S$ increase (which is the opposite when $S$ is small), and that Byzantine nodes follow the algorithm (except when training). Hence, corrupted parameters (that interacts with the trigger pattern) are less likely to be updated by honest nodes. However, Byzantine nodes successfully introduce (albeit not as high as they want) the unwanted behavior, without affecting drastically the normal classification task, as we compare the results of the different result of $S$ studied and the baseline on the right column of Figure~\ref{fig:random-churn-free}: Byzantine nodes induce from 10 to 20 times more misclassification compared to the baseline. Overall, for the 4 topologies studied, namely Erd\H{o}s-Rényi, 20 fan-out, 20 random-regular and Watts-Strogatz,  when Byzantine nodes are placed randomly, against this particular attack, honest nodes are more resilient when the system is using a (very) high number of partitions $S$.

\begin{figure}[htbp]
	\subfloat[Watts-Strogatz]{
		\includegraphics[width=0.5\textwidth]{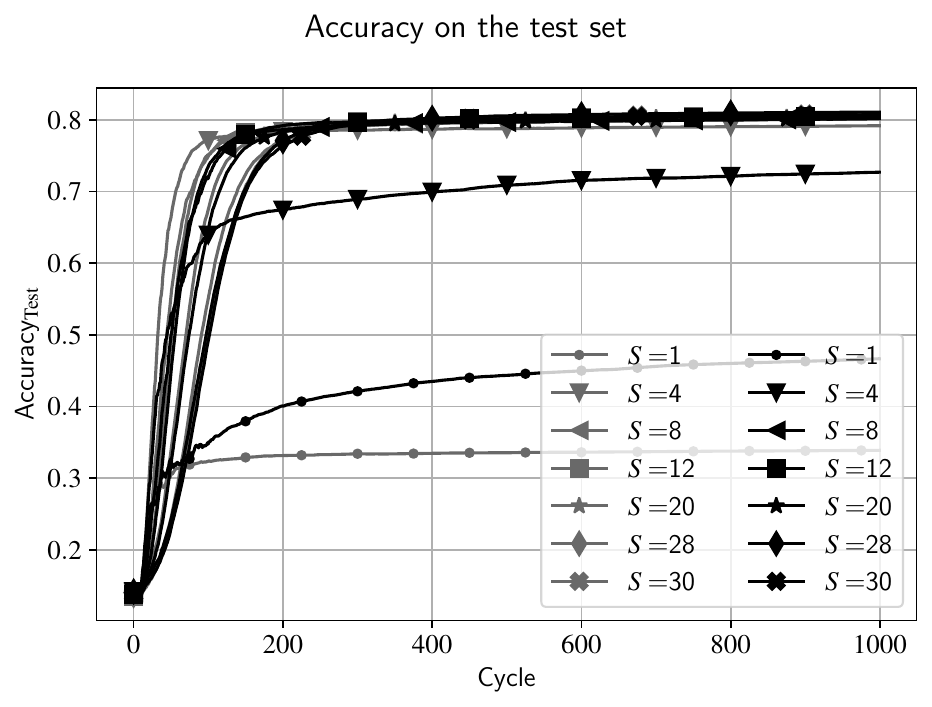}
		\hfill
		\includegraphics[width=0.5\textwidth]{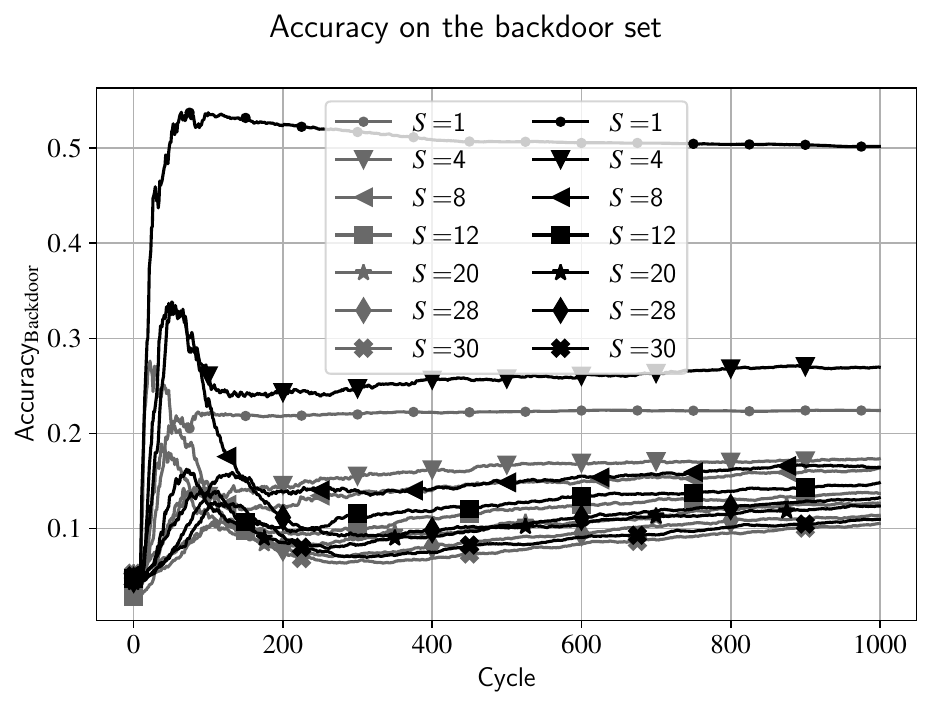}
	}

	\subfloat[Zipf law based degree distribution]{
		\includegraphics[width=0.5\textwidth]{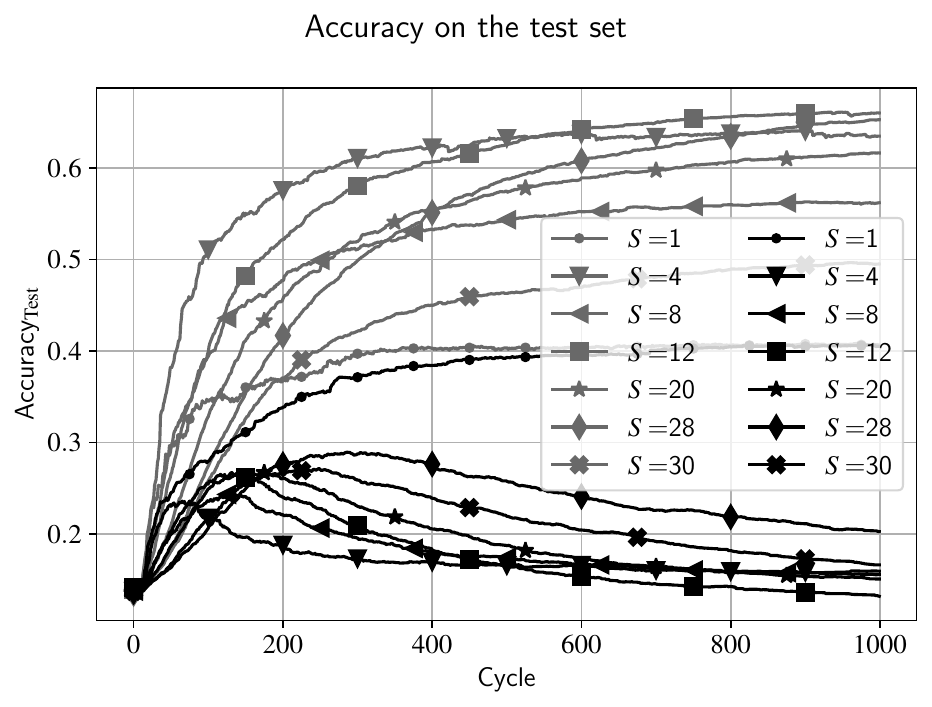}
		\hfill
		\includegraphics[width=0.5\textwidth]{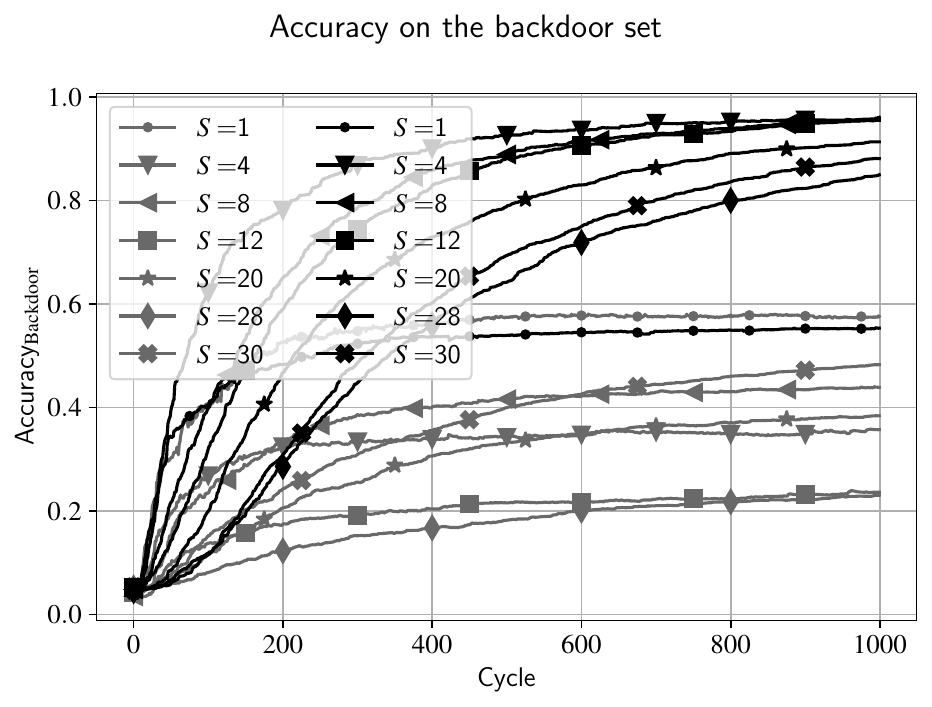}
	}
	\caption{Accuracy on the test set and backdoor set for different topologies with $n = 150$ and $f=45$ with random Byzantine placement strategy (gray) and classical Byzantine placement strategy (black).}
	\label{fig:classical_vs_rd-churn-free}
 \vspace{-0.75cm}
\end{figure}

In Figure~\ref{fig:classical_vs_rd-churn-free}, we compare the \textit{classical} and \textit{random} strategy for Byzantine nodes, in systems that have hubs or where the degree distribution is skewed. Usually, for $S$ fixed, the \textit{classical} strategy is more detrimental for honest nodes than the \textit{random strategy}, especially for the Zipf case or when $S$ is low. On the left column of Figure~\ref{fig:classical_vs_rd-churn-free}, we can see that the accuracy of honest nodes on the test set drop drastically when Byzantines nodes are placed \textit{classically} compared to the \textit{random} strategy. In average, we observe a difference of 6\% and 38\% on the accuracy for the test set, for Watts-Strogatz and Zipf-based topologies  respectively,  which is not what Byzantine nodes want for honest nodes, that do classify the way Byzantine nodes want in the backdoor set (right column of Figure~\ref{fig:classical_vs_rd-churn-free}): in average, we observe a difference of 1\% and 48\%  on the accuracy for the backdoor set between the 2 strategies for Watts-Strogatz and Zipf-based topologies  respectively. 

Considering the case where Byzantine nodes are placed randomly, Watts-Strogatz, and the other three topologies studied previously, are more suitable for honest nodes compared to the Zipf-based topology as they are more resilient against this attack. We notice that for Watts-Strogatz topology, when $S$ is high, that Byzantine placed classically has almost the same impact as if they were placed randomly, this can be attributed to the small-world property of the topology and the fact that the backdoor is too diluted.

 Interestingly, for Zipf-based topology, when Byzantine nodes are placed randomly, excluding $S=1$, on the right column, it sometimes seems detrimental for honest nodes that the system use a high number of partitions $S$ compared to the other 4 topologies already studied previously. 

\noindent
{\bf Churn scenario.}
Here,  we study the case where nodes can disconnect and reconnect as described in Section~\ref{sec:methodology} and have a 20\%  chance to be online.

\begin{figure}[htbp]
	\subfloat{
		\includegraphics[width=0.5\textwidth]{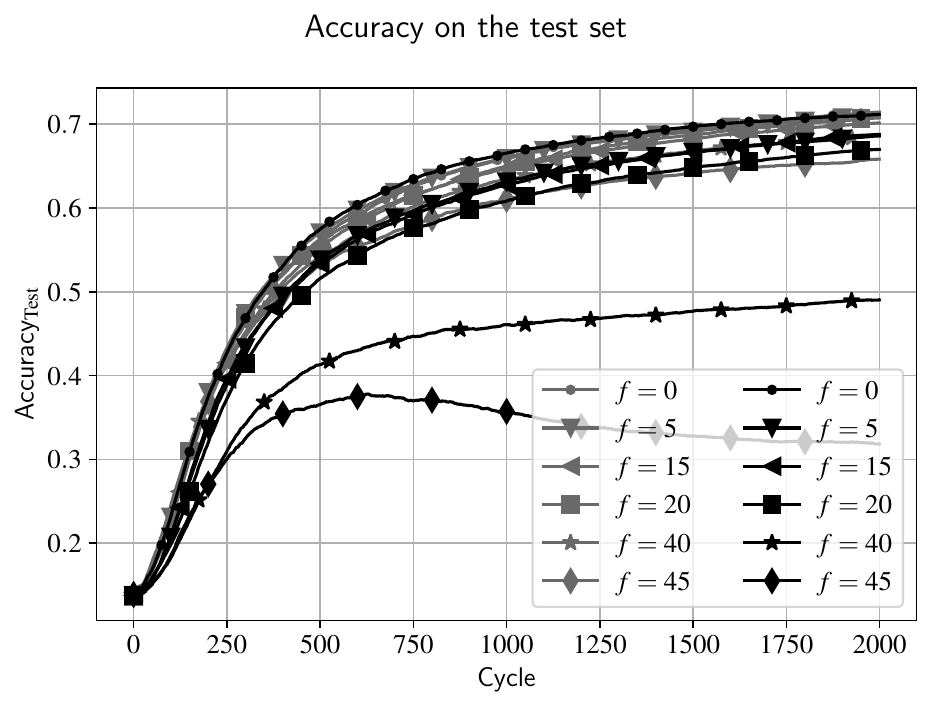}
		\hfill
		\includegraphics[width=0.5\textwidth]{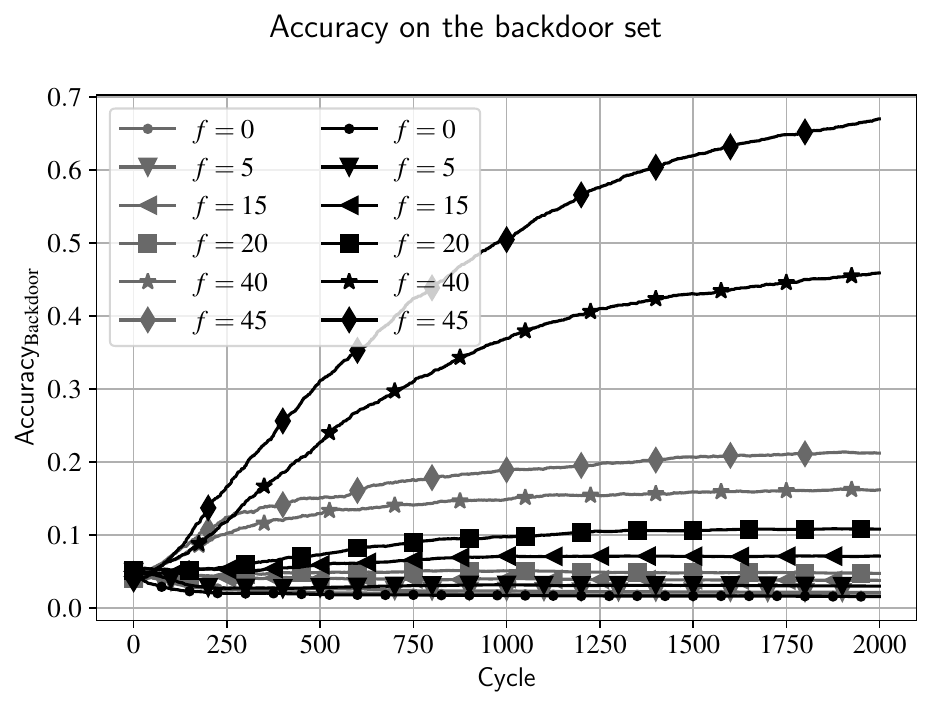}
	}
	\caption{Accuracy on the test and backdoor set for $n=150$, $S=8$ for $f \in \{0, 5, 15, 20, 25, 40, 45\}$ with the random (gray) and classical (black) placement strategy when nodes degree follow a Zipf law.}
	\label{fig:zipf_churn_compare_f_different_strat}
 \vspace{-0.75cm}
\end{figure}

In Figure~\ref{fig:zipf_churn_compare_f_different_strat}, we fix the number of partitions to $S=8$ and number of nodes to $n=150$, and study the effect of Byzantine placement strategy and numbers up to $f=45$ when nodes degree follow a Zipf law distribution. We can see, on the right column of Figure~\ref{fig:zipf_churn_compare_f_different_strat}, that Byzantine nodes, when placed with the \textit{classical} and \textit{random} with $f = 5 $, for the random and classical strategy, it is almost as if there is no attack (case $f=0$). While the \textit{classical} strategy is more harmful to the network for $f$ fixed, considering the backdoor set, we note  the fact that $f=40$ Byzantine nodes selected with the \textit{random} strategy is about 5\% more harmful than $f=20$ Byzantine nodes selected with the \textit{classical} strategy.

\begin{figure}[htbp]
	\subfloat[Erd\H{o}s-Rényi]{
		\includegraphics[width=0.5\textwidth]{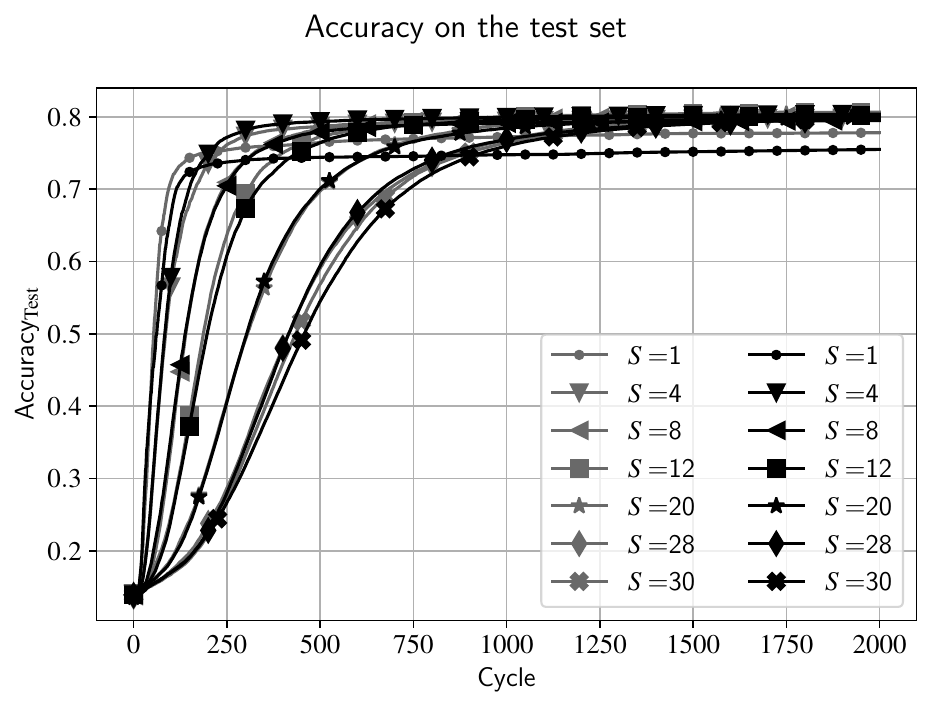}
		\hfill
		\includegraphics[width=0.5\textwidth]{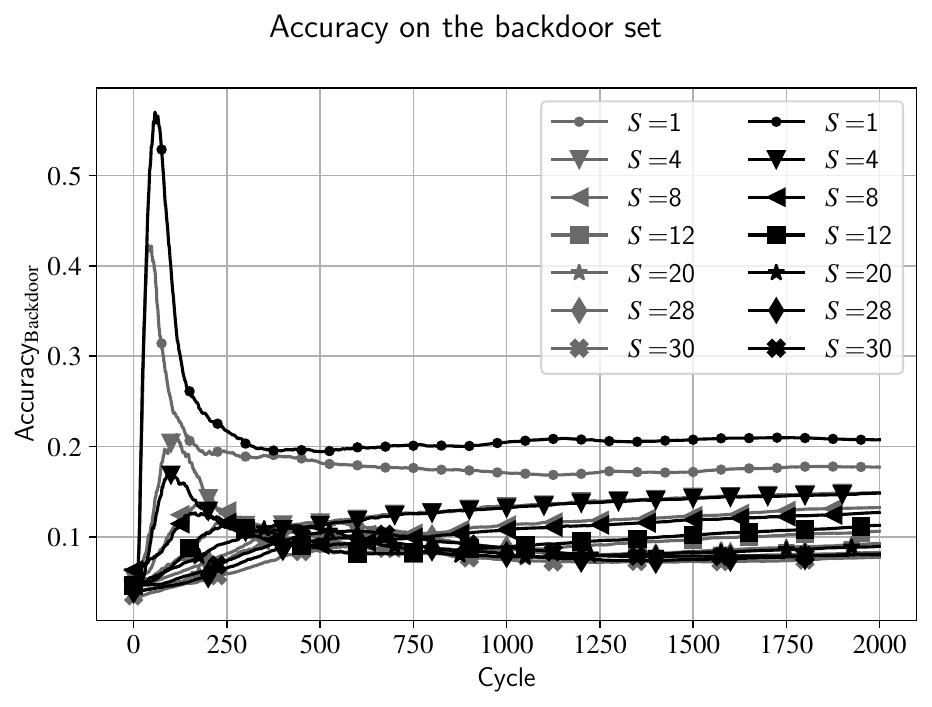}}

	\subfloat[20 fan-out]{
		\includegraphics[width=0.5\textwidth]{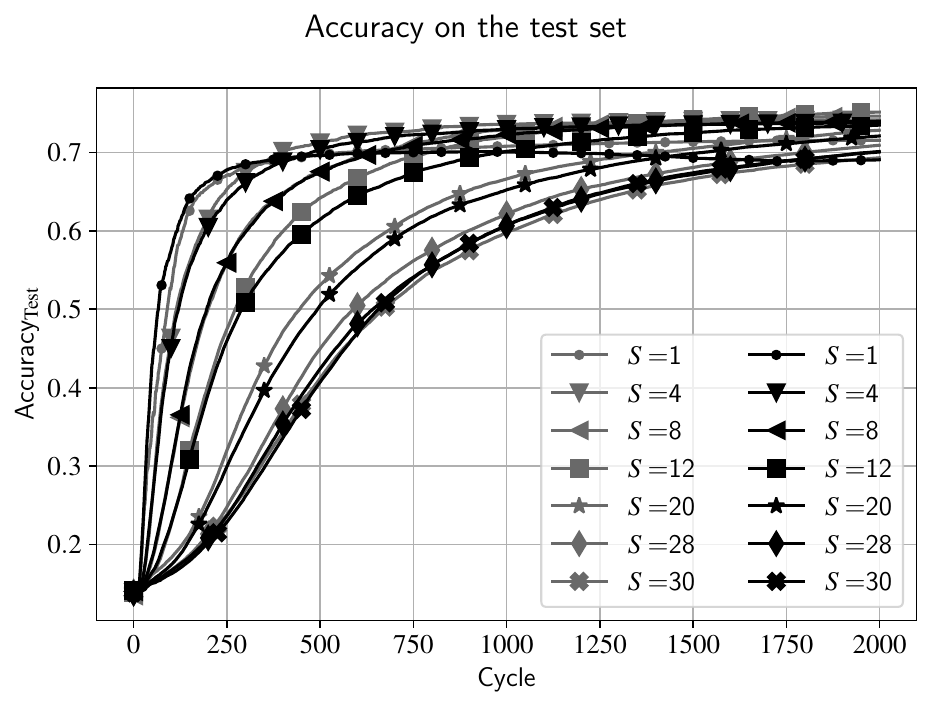}
		\hfill
		\includegraphics[width=0.5\textwidth]{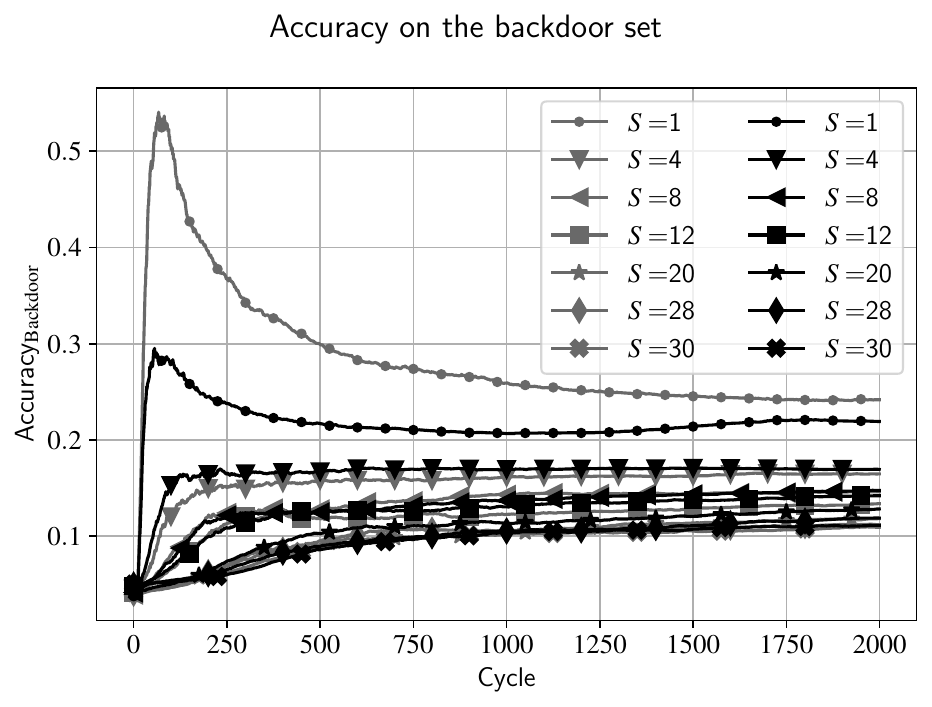}}

	\subfloat[random 20-regular]{
		\includegraphics[width=0.5\textwidth]{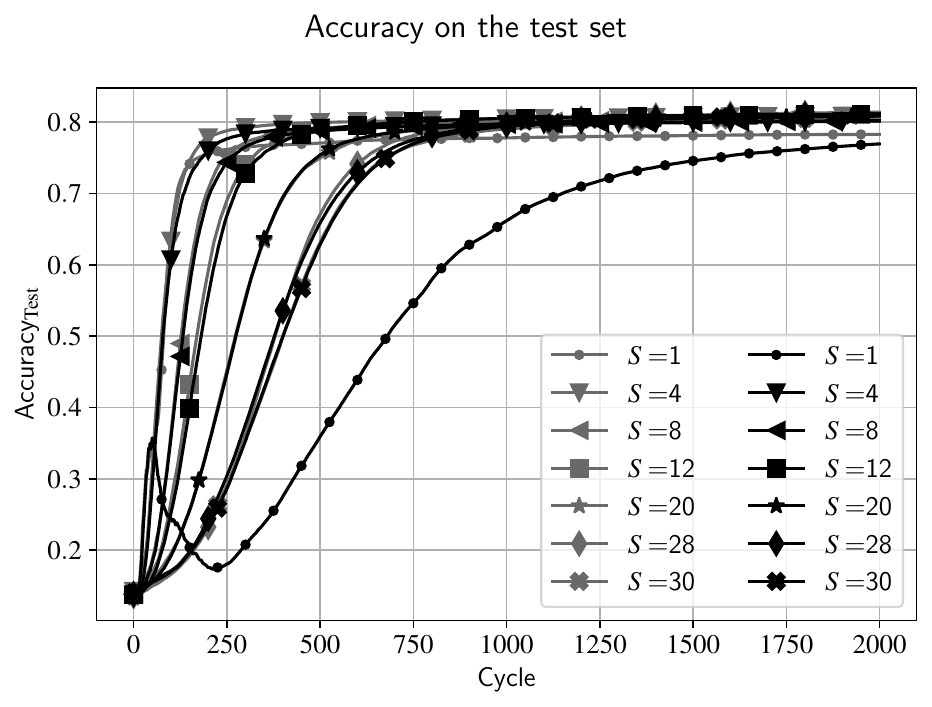}
		\hfill
		\includegraphics[width=0.5\textwidth]{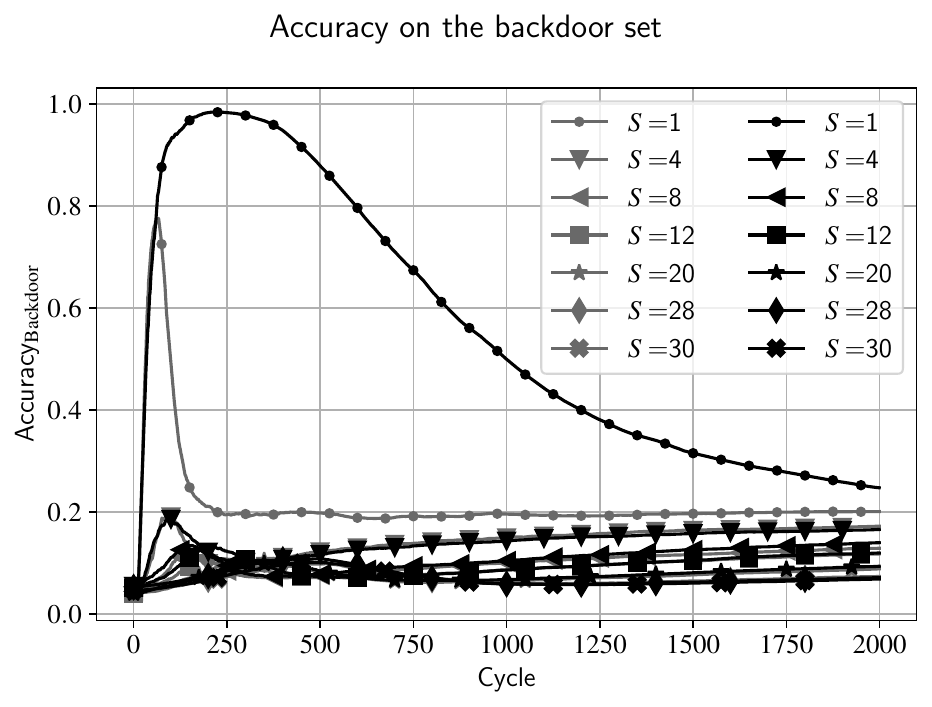}
		}

		\subfloat[Watts-Strogatz]{
			\includegraphics[width=0.5\textwidth]{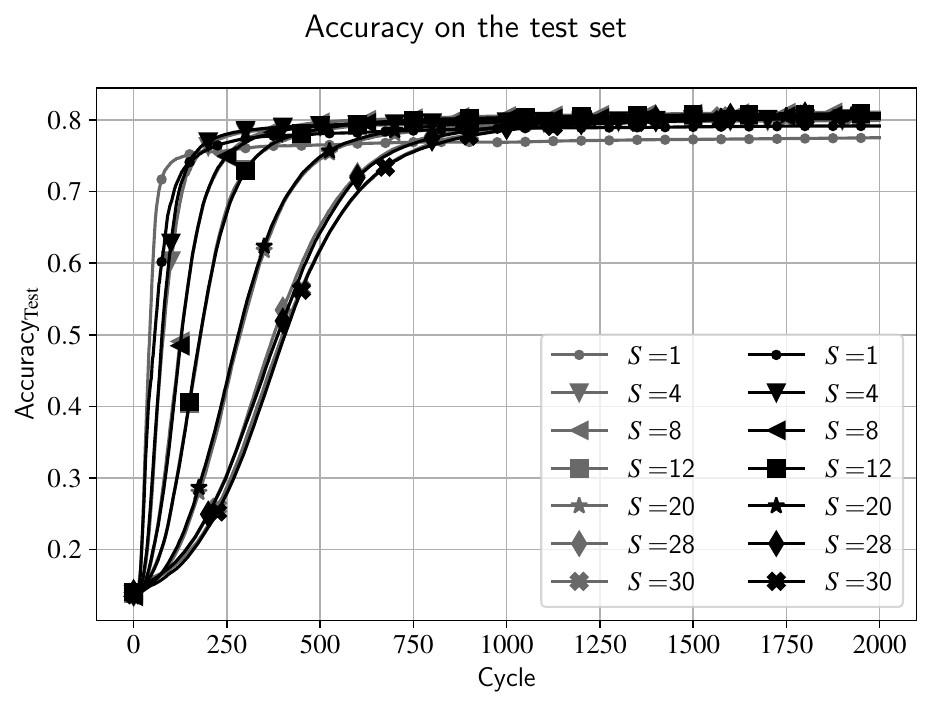}
			\hfill
			\includegraphics[width=0.5\textwidth]{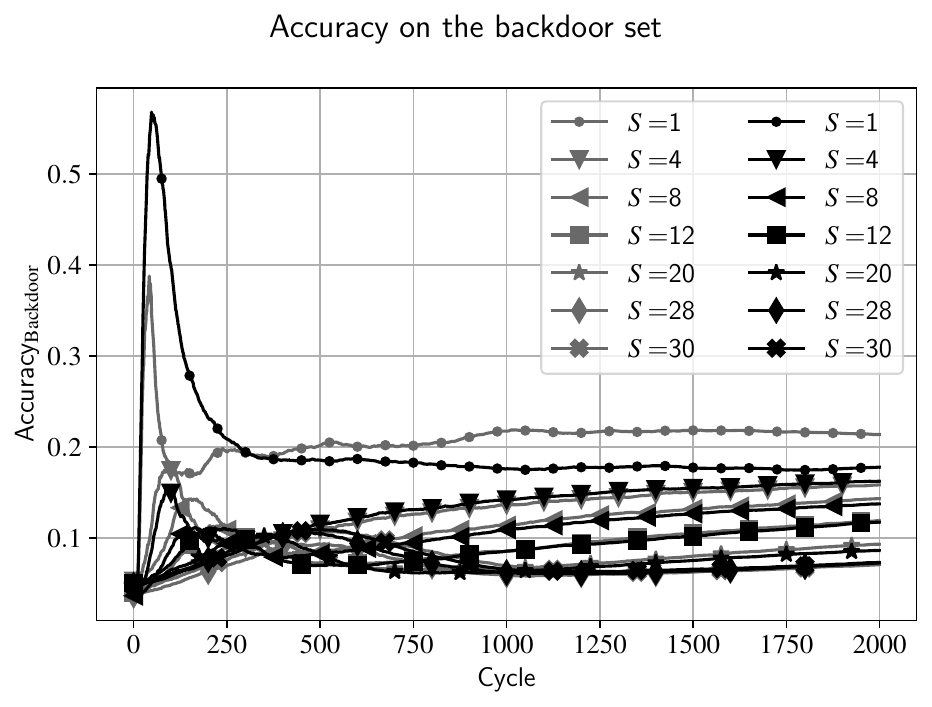}
			}
			\caption{Accuracy on the test set and backdoor set for different topologies with $n = 100 \text{ (gray) and } 150$ (black) with $f=30 \text{ and } 45 \text{ respecively}$.}
			\label{fig:compare_nb_mal-churn}
\end{figure}

In Figure~\ref{fig:compare_nb_mal-churn}, we again study $n=100$ (resp. $n=150$) and $f=30$ (resp. $f=45$), with the \textit{random strategy}, similarly to Figure~\ref{fig:random-churn-free}. We see that honest nodes perform as well as in the \textit{churn-free} scenario on the test set (left column of Figure~\ref{fig:compare_nb_mal-churn}), still due to the fact that data is distributed in an i.i.d. fashion. In this scenario, honest nodes are also slower compared to the churn-free scenario, which is also expected as they are offline most of the time, hence, they are less learning exchanges between nodes for a given number of rounds compared to the churn-free scenario.

Interestingly, while $S=1$ is again a special case here, unlike the churn-free scenario, honest nodes classify badly clean inputs at the beginning, but they converge to a much better solution at the end, as they successfully manage to classify clean inputs with a better success rate and are less affected by the attack compared to the churn-free scenario. This is due to the fact that all nodes are most of the time offline, Byzantine nodes included, hence, the attack is not as powerful as in the churn-free scenario, but we notice, that the results are close with the other value of $S$ studied.
In the journal version of this work, we are planning to study the churn scenario with the assumption that Byzantines nodes are always online.
Overall, we notice that values presented in the churn-free scenario (Figure~\ref{fig:random-churn-free}) and the churn scenario (Figure~\ref{fig:compare_nb_mal-churn}) are relatively close. This might be explained by the fact that the churn is probabilistic, combining with the fact that the choice of the $i^{\text{th}}$ to send is random, hence, over time, the churn-free scenario, looks very similar to the churn-free case (except when $S=1$).

\section{Conclusion}
\label{sec:ccl}
\vspace{-0.5cm}
In this paper, we proposed a methodology to study the resilience of gossip learning in the presence of poisoning attacks. As case study we target the compression mechanism of GL algorithm proposed by Heged\H{u}s et al.~\cite{hegedus.etal_feb2021}. 
We investigate its resilience in a broad range of  topologies (e.g. Erd\H{o}s-Rényi, 20 fan-out, random 20-regular, Watts-Strogatz and Zipf-based graph) in both churn-free and churn scenarios. Our findings show that the communication optimizations and the choice of the underlying topology are not always favorable to the honest nodes. Moreover, the distribution of the Byzantine nodes has also a strong impact on the accuracy metric.

Usually, in the churn-free and churn cases, when Byzantine nodes are placed randomly, the use of a low number of partitions (i.e. bigger messages), is usually detrimental for honest nodes. This is not necessarily true if the topology is Zipf-based, in the churn-free scenario.

In the churn-free scenario, when Byzantine nodes are placed using the \textit{classical} strategy in Watts-Strogatz and Zipf topologies, the attacks can exhibit a completely different impact on the network compared to when they are placed randomly. We observe that in average, Byzantine nodes cause a drop of 1\% and 38\% on the usual classification task and a 6\% and 48\% increase on the backdoor task respectively.

For honest nodes, when Byzantine nodes are placed randomly, we do not observe noticeable differences between the churn-free and churn scenario (except when nodes use bigger messages, as they are more resilient against the attack in the latter scenario), this might be attributed to the probabilistic churn and probabilistic nature of the GL algorithm.

In the future we plan to extend this work to more complex datasets and data distribution, and study other proposals for GL algorithm including some that take into account the possibility of poisoning data and models.

\bibliographystyle{abbrv} 
\bibliography{Biblio_BibTeX.bib}

\end{document}